# Does the Helmholtz theorem of vector decomposition apply to the wave fields of electromagnetic radiation?*

## A. M. Stewart

*Emeritus Faculty, The Australian National University, Canberra, ACT 0200, Australia.*
*http://grapevine.net.au/~a-stewart/index.html*

**Abstract**
The derivation of the Helmholtz theorem of vector decomposition of a 3-vector field requires that the field satisfy certain convergence properties at spatial infinity. This paper investigates if time-dependent electromagnetic radiation wave fields of point sources, which are of long range, satisfy these requirements. It is found that the requirements are satisfied because the fields give rise to integrals over the radial distance $r$ of integrands of the form $\sin(kr)/r$ and $\cos(kr)/r$. These Dirichlet integrals converge at infinity as required.

## 1. Introduction

The Helmholtz theorem of vector field decomposition [1-16] states that a three-vector field $\mathbf{F}(\mathbf{r},t)$ (where $\mathbf{r}$ contains the three spatial coordinates and $t$ is the time) that vanishes at spatial infinity can, under certain conditions, be expressed as the sum of a gradient and a curl

$$\mathbf{F}(\mathbf{r},t) = -\nabla \frac{1}{4\pi} \int d^3 r' \frac{\nabla' \cdot \mathbf{F}(\mathbf{r}',t)}{|\mathbf{r}-\mathbf{r}'|} + \nabla \times \frac{1}{4\pi} \int d^3 r' \frac{\nabla' \times \mathbf{F}(\mathbf{r}',t)}{|\mathbf{r}-\mathbf{r}'|} \quad (1)$$

where $\nabla$ is the gradient with respect to $\mathbf{r}$ ($\nabla'$ is the gradient with respect to $\mathbf{r}'$). Although the theorem is invariably derived for time-independent fields [1,2,4], it is natural to ask if it can be extended to time-dependent ones [5,16-19]. There have been two approaches to this question. In the first [16-18,20], the kernels of the integrals are evaluated at a retarded time. To do this a time derivative term needs to be added to (1) and it is clear that assumptions about the field equations that generate the time dependence of the field have to be made. In the second method [3,6-9,11,12] the time parameter $t$ is simply included in (1) as shown. This is valid because, as noted in the next section, the derivation of (1) involves only spatial derivatives and these carry through as before if the time parameter does not change. No assumption is made about the equations that generate the field. This method had been used to construct electromagnetic potentials of relevance to the Aharonov-Bohm effect [12]. The two approaches have been compared by Woodside [14].

　　　　However, the validity of the Helmholtz theorem of the decomposition of vector fields (1) holds only if the field possesses certain convergence properties at infinity and elsewhere. The electromagnetic radiation wave fields $\mathbf{E}$ and $\mathbf{B}$ generated by the motion of charges are time-dependent and have leading terms of long range, which decay slowly at large distances. This paper investigates if they converge fast enough for (1) to be valid. In section 2 we review the derivation of the Helmholtz decomposition (1) and the convergence properties that are required. In section 3 we reconstruct the derivation of the fields for harmonically varying electric dipole and magnetic dipole radiation as well as those of static fields. In section 4 we investigate if these fields converge fast enough for (1) to be valid and find that they do (but only just). In section 5 it is confirmed that two surface integrals required to obtain expressions for the potentials in terms of their fields [6,7,12] do indeed vanish as needed. In section 6 we recall that the vector potential can be subjected to an arbitrary gauge transformation and consider the conditions that the transformation needs to satisfy for the Helmholtz decomposition to be valid for that field. We find that electric dipole radiation of a point source can be expressed in the Helmholtz form both in the Lorenz and Coulomb [21,22] gauges. In section 7 the results of the paper are summarised. In the appendix we report a situation in which the boundary integrals do not vanish.

**2. Convergence requirements of the Helmholtz theorem**
The validity of (1) is confirmed either by constructing it from equation (2) [11] or, more usually, by taking its divergence and curl and checking for consistency. The latter method is quicker [2] and we adopt it here. First we take the divergence of (1). The second term vanishes because the divergence of a curl is zero. The first term reproduces div**F** (∇·**F**) by making use of the standard identity [23]

$$\delta(\mathbf{r}) = -(1/4\pi)\nabla^2(1/r) \tag{2}$$

where $r$ is the magnitude $|\mathbf{r}|$ of the vector **r**. Next, we take the curl of (1). The first term vanishes, and using the vector identity

$$\nabla \times \nabla \times \mathbf{W} = -\nabla^2 \mathbf{W} + \nabla(\nabla \cdot \mathbf{W}) \tag{3}$$

with

$$\mathbf{W}(\mathbf{r}) = \frac{1}{4\pi} \int d^3 r' \frac{\nabla' \times \mathbf{F}(\mathbf{r}')}{|\mathbf{r} - \mathbf{r}'|} \tag{4}$$

it is seen, using (2), that the first term of the right hand side of (3) reproduces curl**F** (∇x**F**). We need the second term of (3) to vanish and for this it is sufficient that (∇·**W**) is zero.

Using the relations

$$\nabla \cdot [\frac{\nabla' \times \mathbf{F}(\mathbf{r}')}{|\mathbf{r} - \mathbf{r}'|}] = \nabla' \times \mathbf{F}(\mathbf{r}') \cdot \nabla \frac{1}{|\mathbf{r} - \mathbf{r}'|}$$
$$= -\nabla' \times \mathbf{F}(\mathbf{r}') \cdot \nabla' \frac{1}{|\mathbf{r} - \mathbf{r}'|} \tag{5}$$

and

$$\nabla' \cdot [\frac{\nabla' \times \mathbf{F}(\mathbf{r}')}{|\mathbf{r} - \mathbf{r}'|}] = \nabla' \times \mathbf{F}(\mathbf{r}') \cdot \nabla' \frac{1}{|\mathbf{r} - \mathbf{r}'|} + \frac{\nabla' \cdot \nabla' \times \mathbf{F}(\mathbf{r}')}{|\mathbf{r} - \mathbf{r}'|} \tag{6}$$

and, remembering that the divergence of a curl vanishes, we can transform div**W** into a surface integral evaluated at infinity

$$\nabla \cdot \mathbf{W}(\mathbf{r}) = -\frac{1}{4\pi} \int d^3 r' \nabla' \cdot [\frac{\nabla' \times \mathbf{F}(\mathbf{r}')}{|\mathbf{r} - \mathbf{r}'|}]$$
$$= -\frac{1}{4\pi} \oiint_\infty d\mathbf{S}' \cdot [\frac{\nabla' \times \mathbf{F}(\mathbf{r}')}{|\mathbf{r} - \mathbf{r}'|}] \tag{7}$$

and, as *r'* approaches infinity and *r* remains finite,

$$\nabla \cdot \mathbf{W}(\mathbf{r}) = -\frac{1}{4\pi} \oiint_\infty d\Omega' r'\hat{\mathbf{r}}' \cdot [\nabla' \times \mathbf{F}(\mathbf{r}')] \tag{8}$$

where spherical coordinates {*r'*, *θ'*, *φ'*} and vector components are used throughout the paper. The surface element is given by $d\mathbf{S}' = \hat{\mathbf{r}}' r'^2 d\Omega'$, where d**S**' is the vector surface area on the sphere of radius *r'* directed in the direction of **r**' and d$\Omega$ is the solid angle. If the integral (8) vanishes, the convergence condition will be satisfied. This requires the radial component of curl**F** to vanish at infinity faster than 1/*r'*.

However, each term in (1) must be finite. The first term is

$$-\nabla \frac{1}{4\pi} \int d^3 r' \frac{\nabla' \cdot \mathbf{F}(\mathbf{r}')}{|\mathbf{r} - \mathbf{r}'|} = -\frac{1}{4\pi} \int d^3 r' [\nabla' \cdot \mathbf{F}(\mathbf{r}')] \nabla \frac{1}{|\mathbf{r} - \mathbf{r}'|} \tag{9}$$

Apart from at singularities of **F**, places where the integral over **r**' becomes dangerous are at **r**' → **r** and **r**' → ∞. For the first we use **s** = **r**' - **r** to get from (9) as **s** → 0

$$-\frac{\nabla \cdot \mathbf{F}(\mathbf{r})}{4\pi} \int_0 d^3 s \frac{\mathbf{s}}{|\mathbf{s}|^3} = -\frac{\nabla \cdot \mathbf{F}(\mathbf{r})}{4\pi} \int d\Omega_s \int_0 s^2 ds \frac{\mathbf{s}}{|\mathbf{s}|^3}$$

$$= -\frac{\nabla \cdot \mathbf{F}(\mathbf{r})}{4\pi} \int d\Omega_s \hat{\mathbf{s}} \int_0 ds \tag{10}$$

The integral of the unit vector over solid angle gives zero so this term gives no problems. If div**F** is expanded about **r**' = **r** in a Taylor series, terms proportional to components of **s** will appear in the numerator and will give no divergent behaviour.

As *r'* goes to infinity, after taking the gradient, the integral (9) becomes

$$-\frac{1}{4\pi} \int d^3 r' \nabla' \cdot \mathbf{F}(\mathbf{r}') \frac{\hat{\mathbf{r}}'}{|\mathbf{r}'|^2} = -\frac{1}{4\pi} \int d\Omega' \int^\infty dr' [\nabla' \cdot \mathbf{F}(\mathbf{r}')] \hat{\mathbf{r}}' \tag{11}$$

For this integral to converge at infinity it suffices that $\nabla' \cdot \mathbf{F}(\mathbf{r}')$ vanishes at infinity faster than 1/*r'* for all *θ'*, *φ'*. Next we consider the second term of (1)

$$\frac{1}{4\pi} \int d^3 r' \nabla \times \frac{\nabla' \times \mathbf{F}(\mathbf{r}')}{|\mathbf{r} - \mathbf{r}'|} = \frac{-1}{4\pi} \int d^3 r' [\nabla' \times \mathbf{F}(\mathbf{r}')] \times \nabla \frac{1}{|\mathbf{r} - \mathbf{r}'|} \tag{12}$$

The argument for the behaviour at **r**' → **r** goes as before. Performing the gradient operation and letting **r**' approach infinity (12) becomes

$$\frac{1}{4\pi} \int d\Omega' \int^\infty dr' [\nabla' \times \mathbf{F}(\mathbf{r}')] \times \hat{\mathbf{r}}' \tag{13}$$

For this integral to converge at infinity it suffices that all the components of $[\nabla' \times \mathbf{F}(\mathbf{r}')] \times \hat{\mathbf{r}}$ vanish at infinity faster than $1/r'$.

**3. Electromagnetic fields**

We consider a system of harmonically moving charges close to the origin of coordinates with total charge $q$ and total magnetic moment $\mathbf{m}$, and we find the potentials and fields at a large distance from the charges. The gauge of the scalar potential V and vector potential $\mathbf{A}$ is chosen to satisfy the Lorenz gauge condition [2]

$$\nabla \cdot \mathbf{A}(\mathbf{r}) + \frac{1}{c^2}\frac{\partial V}{\partial t} = 0 \quad . \tag{14}$$

and give rise to the four source-free Maxwell equations.

First we consider static fields with potentials obtained from the electric sources in the standard way [2,24]

$$V(\mathbf{r}) = q/(4\pi\varepsilon_0 r) \tag{15}$$

$$\begin{aligned}\mathbf{A}(\mathbf{r}) &= \mu_0 \mathbf{m} \times \mathbf{r}/4\pi r^3 \\ &= (\mu_0 m/4\pi r^2)\{0, 0, \sin\theta\}\end{aligned} \tag{16}$$

with a charge $q$ and a magnetic dipole $\mathbf{m}$ directed along the $z$-axis at the origin. These potentials also satisfy the Coulomb gauge condition $\nabla \cdot \mathbf{A}(\mathbf{r}) = 0$. From them we obtain

$$\begin{aligned}\mathbf{B} &= \nabla \times \mathbf{A} \\ &= (\mu_0 m/4\pi r^3)\{2\cos\theta, \sin\theta, 0\}\end{aligned} \tag{17}$$

$$\begin{aligned}\mathbf{E} &= -\nabla V - \partial \mathbf{A}/\partial t \\ &= \frac{q}{4\pi\varepsilon_0 r^2}\{1, 0, 0\}\end{aligned} \tag{18}$$

with div$\mathbf{B}$ = 0, curl$\mathbf{B}$ = 0, div$\mathbf{E}$ = 0, curl$\mathbf{E}$ = 0, and

$$(\nabla \times \mathbf{A}) \times \hat{\mathbf{r}} = (\mu_0 m/4\pi r^3)\{0, 0, -\sin\theta\} \quad . \tag{19}$$

Next we consider the coordinate dependence of electric dipole radiation fields. The potentials go as [2]

$$V(\mathbf{r},t) = (\frac{p\mu_0}{4\pi})c^2(\frac{\cos\theta}{r})\{-(\frac{\omega}{c})\sin[\omega(t-r/c)] + \cos[\omega(t-r/c)]/r\} \quad , \tag{20}$$

$$\mathbf{A}(\mathbf{r},t) = (\frac{p\mu_0}{4\pi})(\frac{\omega}{r})\sin[\omega(t-r/c)]\{-\cos\theta, \sin\theta, 0\} \quad , \tag{21}$$

with the oscillating electric dipole moment $p\sin[\omega t]$ directed along the *z*-axis. We drop the terms in the first brackets in (20-21) for subsequent brevity to get

$$\nabla \cdot \mathbf{A}(\mathbf{r}) = \frac{\omega^2 \cos\theta}{cr}\cos[\omega(t - r/c)] + \frac{\omega\cos\theta}{r^2}\sin[\omega(t - r/c)] \quad . \quad (22)$$

The fields, displaying the leading terms in 1/*r*, are

$$\mathbf{B} = \{0, 0, -(\frac{\omega^2 \sin\theta}{cr})\cos[\omega(t - r/c)] - (\frac{\omega\sin\theta}{r^2})\sin[\omega(t - r/c)]\} \quad (23)$$

$$\mathbf{E} = \{-(\frac{2\omega c \cos\theta}{r^2})\sin[\omega(t - r/c)], -(\frac{\omega^2 \sin\theta}{r})\cos[\omega(t - r/c)], 0\} \quad (24)$$

from which we obtain the Poynting vector [2] $\mathbf{P} = \mathbf{E}\times\mathbf{B}/\mu_0$

$$\mathbf{P} \approx \{\omega^4 \sin^2\theta \cos^2[\omega(t - r/c)]/c^3 r^2, O[r^{-3}], 0\} \quad . \quad (25)$$

Other relations needed are

$$\nabla \times \mathbf{E} = \{0, 0, (\frac{\omega^2 \sin\theta}{r^2})\cos[\omega(t - r/c)] - (\frac{\omega^3 \sin\theta}{cr})\sin[\omega(t - r/c)]\} \quad (26)$$

$$\nabla \times \mathbf{B} = \{-\frac{2\omega^2 \cos\theta}{cr^2}\cos[\omega(t - r/c)], \frac{\omega^3 \sin\theta}{c^2 r}\sin[\omega(t - r/c)], 0\} \quad (27)$$

and hence the expressions needed for use in (13)

$$(\nabla \times \mathbf{A}) \times \hat{\mathbf{r}} = \{0, -(\frac{\omega^2 \sin\theta}{cr})\cos[\omega(t - r/c)] - (\frac{\omega\sin\theta}{r^2})\sin[\omega(t - r/c)], 0\} \quad , \quad (28)$$

$$(\nabla \times \mathbf{E}) \times \hat{\mathbf{r}} = \{0, (\frac{\omega^2 \sin\theta}{r^2})\cos[\omega(t - r/c)] - (\frac{\omega^3 \sin\theta}{cr})\sin[\omega(t - r/c)], 0\} \quad , \quad (29)$$

$$(\nabla \times \mathbf{B}) \times \hat{\mathbf{r}} = \{0, 0, \frac{\omega^2 \sin\theta}{cr^2}\cos[\omega(t - r/c)] - \frac{\omega^3 \sin\theta}{c^2 r}\sin[\omega(t - r/c)]\} \quad . \quad (30)$$

Lastly, we consider magnetic dipole radiation [2]. With the oscillating magnetic dipole *m* directed along the *z*-axis, this gives rise to zero scalar potential and vector potential

$$\mathbf{A} = (\frac{m\mu_0}{4\pi})\frac{\sin\theta}{r}(\cos[\omega(t - r/c)]/r - (\omega/c)\sin[\omega(t - r/c)])\{0, 0, 1\} \quad . \quad (31)$$

We drop the term in the first brackets of (31) for brevity. The divergence of **A** is zero. The fields are

$$\mathbf{B} = \{-(\frac{2\omega\cos\theta}{cr^2})\sin[\omega(t - r/c)], -(\frac{\omega^2 \sin\theta}{c^2 r})\cos[\omega(t - r/c)], 0\} \quad , \quad (32)$$

$$\mathbf{E} = \{0, 0, (\frac{\omega^2 \sin\theta}{cr})\cos[\omega(t - r/c)]\} \quad , \quad (33)$$

$$\mathbf{P} \approx \{\omega^4 \sin^2\theta \cos^2[\omega(t - r/c)]/c^3 r^2, O[r^{-3}], 0\} \quad . \quad (34)$$

$$\nabla \times \mathbf{E} = \{(\frac{2\omega^2}{cr^2})\cos\theta\cos[\omega(t - r/c)], -(\frac{\omega^3}{c^2 r})\sin\theta\sin[\omega(t - r/c)], 0\} \quad (35)$$

$$\nabla \times \mathbf{B} = \{0, 0, -\frac{\omega^3 \sin\theta}{c^3 r}\sin[\omega(t - r/c)]\} \quad (36)$$

The expressions needed for equation (13) are

$$(\nabla \times \mathbf{A}) \times \hat{\mathbf{r}} = \{0, 0, (\frac{\omega\sin\theta}{cr^2})\sin[\omega(t - r/c)] + (\frac{\omega^2 \sin\theta}{c^2 r})\cos[\omega(t - r/c)]\} \quad , \quad (37)$$

$$(\nabla \times \mathbf{E}) \times \hat{\mathbf{r}} = \{0, 0, -(\frac{\omega^2}{cr^2})\sin\theta\cos[\omega(t - r/c)] + (\frac{\omega^3}{c^2 r})\sin\theta\sin[\omega(t - r/c)]\} \quad , \quad (38)$$

$$(\nabla \times \mathbf{B}) \times \hat{\mathbf{r}} = \{0, -\frac{\omega^3 \sin\theta}{c^3 r}\sin[\omega(t - r/c)] + \frac{\omega^2 \sin\theta}{c^2 r^2}\cos[\omega(t - r/c)], 0\} \quad . \quad (39)$$

None of the potentials or fields depends on the azimuthal coordinate $\phi$.†

**4 Convergence of the Helmholtz terms**
For (1) to be a meaningful representation of the vector field **F** the following conditions must be met: the surface integral of (8) must vanish and the integrals (11) and (13) must converge at infinity.

For static fields the only term that at first sight is not zero for (8), substituting **F** by **A**, is (17). However this term goes as $1/r^3$ and therefore (8) approaches zero at infinity. All three terms of (11) for **A**, **E**, **B** are zero. Equation (13) using **F** = **A** (19) gives

$$\frac{\mu_0 m}{(4\pi)^2} \int_0^{2\pi} d\phi' \int_0^\pi d\theta' \sin\theta' \int^\infty \frac{dr'}{r'^3}\{0, 0, -\sin\theta'\} \quad . \quad (40)$$

Since the integrand does not depend on $\phi'$ the integral over $\phi'$ gives zero when the $\phi'$ component of the vector is projected on the $x$ and $y$ axes. The integral over $\theta'$ is $-\pi/2$ and the integral over $r'$, being the integral of $1/r'^3$, converges as $r' \to \infty$. Therefore the integral vanishes and so static electromagnetic fields may be described by the Helmholtz theorem.

For electric dipole radiation (8) is satisfied for **F** substituted by **A** (23) and by **E** (26) because the scalar product in (8) gives zero. Using **F** = **B** it gives from (27)

$$\nabla \cdot \mathbf{W}(\mathbf{r}) = \frac{\omega^2}{2\pi c} \oiint_\infty d\Omega' \cos\theta \cos[\omega(t - r/c)]/r' \qquad (41)$$

The integral over $\phi'$ is finite and equal to $2\pi$, the integral over $\theta'$, being the integral over $\theta'$ from 0 to $\pi$ of $\sin\theta'\cos\theta'$, vanishes and the term in the integrand that depends on $r'$ vanishes as $r' \to \infty$. Accordingly div**W** vanishes. Next we consider the condition that (11) should converge with **F** substituted by **A** (22). This requires the convergence of the integral

$$\frac{\omega^2}{c} \int_0^{2\pi} d\phi' \int_0^\pi d\theta' \sin\theta' \cos\theta' \int_0^\infty dr' \frac{\cos[\omega(t - r'/c)]}{r'} \hat{\mathbf{r}}' \qquad (42)$$

The integral over $\phi'$ gives $2\pi\cos\theta'$ with a vector direction along the $z$ ($\theta = 0$) axis. The integral of $\sin\theta' \cos^2\theta'$ over $\theta'$ gives $2/3$ and we are left with the radial integral

$$\int_0^\infty dr' \frac{\cos[\omega(t - r'/c)]}{r'} = \cos(\omega t) \int_0^\infty dr' \frac{\cos(kr')}{r'} + \sin(\omega t) \int_0^\infty dr' \frac{\sin(kr')}{r'} \qquad (43)$$

where $k = \omega/c$. The integrals over $r'$ are Dirichlet integrals which both converge at infinity [25]. The sine integrand converges at zero argument also and gives a finite integral

$$\int_0^\infty dx \frac{\sin(kx)}{x} = \frac{\pi}{2} \qquad (44)$$

for $k > 0$ (for $k < 0$ the integral is $-\pi/2$, for $k = 0$ it is zero [25]). Convergence condition (11) is therefore satisfied for **A**. We note that convergence is only just obtained. In gravitation theory [26] it is also found that fields that go as $\exp(iqr)/r$ lead to convergence but fields that go as $1/r$ do not. The quantities div**E** and div**B** are zero so the convergence condition (11) is satisfied for them also.

Next we consider condition (13) for equations (28-30). The angular integrals give a finite value or zero. The leading term of the radial integrals is a Dirichlet integral that converges as before. So (13) is satisfied and electric dipole radiation may be described by (1).

The situation is similar for magnetic dipole radiation. The divergence of **W** of (8) vanishes for **A**, **E** and **B** because the radial components of their curls go to zero at least as fast as $1/r'^2$. The divergence of all three vector fields **A**, **E** and **B** is zero so that the condition (11) is satisfied. For equation (13) the angular integrals are finite or zero and the leading terms of the radial integrals are again Dirichlet integrals that converge at infinity. Therefore both electric dipole and magnetic dipole radiation satisfy the convergence conditions required for (1).

**5. Vanishing of two surface integrals**
Next we examine two surface integrals that are needed for the derivation of the forms of the Coulomb gauge potentials expressed in terms of the fields used in [12]. It is possible to express the well-known expression for the scalar potential [2] in the Coulomb gauge

$$V(\mathbf{r},t) = \frac{1}{4\pi\varepsilon_0} \int d^3r' \frac{\nabla' \cdot \mathbf{E}(\mathbf{r}',t)}{|\mathbf{r} - \mathbf{r}'|} \qquad (45)$$

in the form

$$V(\mathbf{r},t) = \nabla \cdot \int d^3r' \frac{\mathbf{E}(\mathbf{r}',t)}{4\pi |\mathbf{r}-\mathbf{r}'|} \quad . \tag{46}$$

This is done by taking the divergence of the expression $\mathbf{E}(\mathbf{r}',t)/|\mathbf{r}-\mathbf{r}'|$ separately with respect to both $\nabla$ and $\nabla'$ and doing a partial integration to get a surface integral that is required to vanish

$$\int d^3r' \nabla' \cdot [\frac{\mathbf{E}(\mathbf{r}',t)}{|\mathbf{r}-\mathbf{r}'|}] = \oiint_\infty d\mathbf{S}' \cdot [\frac{\mathbf{E}(\mathbf{r}',t)}{|\mathbf{r}-\mathbf{r}'|}] \quad . \tag{47}$$

From (18), (24) and (33) we find that the radial component of **E** vanishes at least as fast as $1/r'^2$ so the surface integral (47) vanishes and the derivation of (46) is valid. The convergence of (45) as $r'$ approaches infinity at infinity is assured by div**E** vanishing at infinity for all source-free fields. The pole of (46) at $\mathbf{r}' = \mathbf{r}$ is harmless as shown in section 2.

The validity of the expression for the vector potential [6,7,11]

$$\mathbf{A}(\mathbf{r},t) = \nabla \times \int d^3r' \frac{\mathbf{B}(\mathbf{r}',t)}{4\pi |\mathbf{r}-\mathbf{r}'|} \tag{48}$$

is verified by taking its curl [7] and using (3). The first term of (3) gives $\mathbf{B}(\mathbf{r},t)$, the second term gives

$$\nabla \int d^3r' \nabla \cdot \frac{\mathbf{B}(\mathbf{r}',t)}{4\pi |\mathbf{r}-\mathbf{r}'|} = -\nabla \int d^3r' \nabla' \cdot \frac{\mathbf{B}(\mathbf{r}',t)}{4\pi |\mathbf{r}-\mathbf{r}'|} \quad . \tag{49}$$

This becomes the surface integral [7]

$$-\nabla \oiint_\infty d\mathbf{S}' \cdot [\frac{\mathbf{B}(\mathbf{r}',t)}{|\mathbf{r}-\mathbf{r}'|}] \quad . \tag{50}$$

From (17), (23) and (32) we find that the radial component of **B** vanishes at least as fast as $1/r'^2$ so the surface integral (50) vanishes and expression (48) is valid. The convergence of (48) as $r'$ approaches infinity obtains for the radiation fields, because, after letting the $\nabla$ act on the $1/|\mathbf{r} - \mathbf{r}'|$ factor, a convergent Dirichlet integral results.

## 6. Gauge transformation of the potentials
If the gauge transformation

$$\mathbf{A} \rightarrow \mathbf{A}' = \mathbf{A} + \nabla \chi(\mathbf{r},t) \tag{51}$$

and

$$V \rightarrow V' = V - \frac{\partial}{\partial t}\chi(\mathbf{r},t) \tag{52}$$

is applied to the vector **A** and scalar $V$ potentials, with the scalar gauge function $\chi$, then the observable fields **E** and **B** are unchanged [22,27]. We ask what conditions the gauge function must satisfy for a transformed vector potential **A'** to still satisfy the conditions required by the Helmholtz theorem. Conditions (8) and (13) are automatically satisfied by the change of vector potential $\Delta \mathbf{A} = \nabla \chi(\mathbf{r},t)$ because the curl of a gradient is zero. Condition (11) holds providing that

$$\intد\Omega' \int_0^\infty dr' [\nabla'^2 \chi(\mathbf{r}',t)] \hat{\mathbf{r}}' \tag{53}$$

converges as $r'$ approaches infinity. A gauge function that satisfies the Laplace equation will satisfy this trivially but other functions may do so also. This may be also seen from (1) as its second term is unchanged by a gauge transformation and the first term leads to condition (53).

The gauge condition used for static fields and for magnetic dipole radiation in section 3 satisfies the Coulomb gauge condition $\nabla \cdot \mathbf{A}(\mathbf{r}) = 0$ as well as the Lorenz gauge condition (13). The Coulomb gauge is of special interest because it has a minimum property: namely that the volume integral of $\mathbf{A}^2$ over all space is a minimum for this gauge [10,28]. The potentials of electric dipole radiation (20, 21) satisfy the Lorenz gauge condition and it is interesting to consider what form these potentials take in the Coulomb gauge.

In the Coulomb gauge the scalar potential $V_C$ of electric dipole radiation takes the well-known instantaneous form [6,22]

$$V_C(\mathbf{r},t) = (\frac{p\mu_0}{4\pi})c^2 \frac{\cos\theta}{r^2} \sin(\omega t) \tag{54}$$

where the electric dipole moment is $p\sin(\omega t)$ directed along the $z$-axis. Jackson [22] has shown that the vector potential $\mathbf{A}_C$ *for electromagnetic radiation* has the simple form

$$\mathbf{A}_C(\mathbf{r},t) = \frac{\mu_0}{4\pi} \int d^3r' \frac{\mathbf{J}(\mathbf{r}',t') - \hat{\mathbf{R}}[\hat{\mathbf{R}} \cdot \mathbf{J}(\mathbf{r}',t')]}{R} \tag{55}$$

where $\mathbf{R} = \mathbf{r} - \mathbf{r}'$, $R = |\mathbf{R}|$ and $t'$ is the retarded time $t' = t - R/c$. The first term on the right hand side of (55) gives the vector potential in the Lorenz gauge.

When $\mathbf{A}_C$ is derived from (55) in the standard way [2,24], it comes to

$$\mathbf{A}_C(\mathbf{r},t) = (\frac{p\mu_0}{4\pi})(\frac{\omega}{r}) \sin[\omega(t - r/c)]\{0, \sin\theta, 0\} \tag{56}$$

instead of (21) in the Lorenz gauge. To order $1/r$ (56) satisfies $\nabla \cdot \mathbf{A}_C(\mathbf{r}) = 0$ and so (11) is satisfied. Also, to order $1/r$, the fields **E** and **B** are the same as in (23) and (24) so (8) and (13) vanish at infinity. Equation (56) may be obtained from (21) using the gauge function $\chi = p\mu_0 c \cos\theta \cos[\omega(t - r/c)]/4\pi r$, and the scalar potential that results from it is $V_C = p\mu_0 c^2 \cos\theta \cos[\omega(t - r/c)]/4\pi r^2$, rather than the exact form (54). However, both forms vary as $1/r^2$ and so do not have any effect on **E** to order $1/r$ and so the scalar potential may just as well be set to zero if terms of order $1/r$ only are considered. We find that the vector potential of the radiation fields of point sources can be expressed in the Helmholtz form both in the Lorenz and Coulomb gauges.

## 7. Summary

The derivation of the Helmholtz theorem of decomposition of a vector field requires that the field satisfy certain convergence properties at spatial infinity. The paper has investigated if these requirements are satisfied by electromagnetic radiation wave fields of point sources, which vanish at infinity but are of long range. It is found that the requirements are satisfied *both* for the Lorenz and Coulomb gauges because the fields give rise to integrals over the radial distance *r* of integrands of the form sin($kr$)/$r$ and cos($kr$)/$r$. When integrated to infinity, these Dirichlet integrals just converge. Since, of electromagnetic fields, radiation fields are the most slowly converging at infinity, it follows that the Helmholtz theorem can be applied to the electromagnetic fields generated by point sources. It is also found that two surface integrals needed to derive the electromagnetic potentials in terms of the fields vanish as required.

**Appendix**

This paper has been devoted to showing that the surface integrals of radiation fields at infinity that arise in the Helmholtz decomposition cause no trouble there. In this appendix we examine one situation where the surface integrals diverge. This involves a method [29,8] of decomposing the angular momentum **J** of the electromagnetic field

$$\mathbf{J} = \int d^3 x\, \mathbf{x} \times [\mathbf{E}(\mathbf{x}) \times \mathbf{B}(\mathbf{x})] \tag{A1}$$

into a spin part and an orbital part. We will find that the method does not work for radiation fields because a surface integrals do not vanish.

We review the method of decomposition. By using the relations **B** = ∇x**A** and the vector relation

$$\mathbf{E} \times (\nabla \times \mathbf{A}) = \sum_m E^m \nabla A^m - (\mathbf{E}\cdot\nabla)\mathbf{A} \tag{A2}$$

we express (A1) as

$$\mathbf{J} = \sum_m \int d^3 x\, E^m (\mathbf{x} \times \nabla) A^m - \int d^3 x\, \mathbf{x} \times (\mathbf{E}\cdot\nabla)\mathbf{A} \ . \tag{A3}$$

The first term of (A3) with **x**x∇ corresponds to orbital angular momentum. The second term is to be manipulated into the form of spin angular momentum, which does not depend linearly on **x**. Of course, the decomposition is meaningless if the gauge of **A** is not fixed. The gauge that is invariably chosen in this situation is the Coulomb gauge with ∇**.A** = 0 [6].

The second term of (A3) is treated in the following manner. We construct the vector **V**:

$$\mathbf{V(x)} = \sum_m \frac{\partial}{\partial x^m}(E^m \mathbf{x} \times \mathbf{A}) \quad \text{or} \quad V(\mathbf{x})|^i = \sum_{mjk} \varepsilon^{ijk} \frac{\partial}{\partial x^m}(E^m x^j A^k) \ . \tag{A4}$$

By using the identity

$$\frac{\partial}{\partial x^m}(E^m x^j A^k) = \delta_{jm} E^m A^k + x^j (A^k \frac{\partial E^m}{\partial x^m} + E^m \frac{\partial A^k}{\partial x^m}) \tag{A5}$$

we get

$$V(\mathbf{x})|^i = \sum_{jk} \varepsilon^{ijk} E^j A^k + \sum_{mjk} \varepsilon^{ijk} x^j (A^k \frac{\partial E^m}{\partial x^m} + E^m \frac{\partial A^k}{\partial x^m}) \tag{A6}$$

or

$$\mathbf{V}(\mathbf{x}) = \mathbf{E} \times \mathbf{A} + \mathbf{x} \times \mathbf{A}(\nabla \cdot \mathbf{E}) + \mathbf{x} \times (\mathbf{E} \cdot \nabla)\mathbf{A} \tag{A7}$$

Integrating over $d^3x$ and with the assumption that $\mathbf{N} = \int d^3x \, \mathbf{V}(\mathbf{x}) = 0$ we obtain

$$\mathbf{J} = \int d^3x \, [\sum_m E^m (\mathbf{x} \times \nabla) A^m + \mathbf{E} \times \mathbf{A} + \mathbf{x} \times \mathbf{A}(\nabla \cdot \mathbf{E})] = \mathbf{L} + \mathbf{S} + \mathbf{B} \tag{A8}$$

The first term $\mathbf{L}$ is denoted the orbital term, the second term $\mathbf{S}$ the spin term and the third term $\mathbf{B}$ is non-zero only in the presence of charge [8,9,29].

For (A8) to be valid it must be shown that $\mathbf{N}$, the integral of $\mathbf{V}$ over all space, is zero. Whether it is zero depends on the nature of the fields. We will find that for static fields the integral is zero and (A8) may be valid, but for radiation fields the integral diverges and (A8) is not valid.

From (A4) we require

$$N^i = \sum_m \int dx^m \, dx^s \, dx^t \, \frac{\partial}{\partial x^m} [E^m (\mathbf{x} \times \mathbf{A})|^i)] = 0 \tag{A9}$$

for all $i$. The volume of integration is taken to be a cube of side $2R$ centred about the origin with sides parallel to the $\{x,y,z\}$ axes. We therefore need

$$\iint dy \, dz \, E^x (\mathbf{x} \times \mathbf{A})|^i |_{x=-R}^{x=R} + \iint dz \, dx \, E^y (\mathbf{x} \times \mathbf{A})|^i |_{y=-R}^{y=R} + \iint dx \, dy \, E^z (\mathbf{x} \times \mathbf{A})|^i |_{z=-R}^{z=R} = 0 \tag{A10}$$

for $i = \{x,y,z\}$. In the first term of (A10) the surface integral is evaluated over the y-z square of area $4R^2$ at $x = R$ minus the same integral at $x = -R$.

For a static source distribution with charge $q$ and magnetic moment $\mathbf{m}$ directed along the z-axis the fields are, in spherical coordinates and ignoring irrelevant constants (15,16),

$$\mathbf{E} \approx \{1, 0, 0\} / r^2 \qquad \text{and} \qquad \mathbf{A}(\mathbf{r}) \approx \{0, 0, 1\} \sin\theta / r^2 \tag{A11}$$

When these are put into one of the terms of (A10), the integrals on each face of the cube will go as $1/R$ and therefore vanish. The decomposition (A8) is therefore valid for static fields. If the cubic box is displaced from the origin by a vector $\alpha R\{1,1,1\}$, where $\alpha$ is a small number, the surface integrals are changed by a term proportional to $\alpha^2/R$, which also vanishes as $R \to \infty$, and the result is unchanged.

However, the situation is not so straightforward for the radiation fields of an electric dipole (56), which go as

$$\mathbf{E} \approx -\{0, 1, 0\} c^2 k^2 \sin\theta \cos(kr) / r \quad \text{and} \quad \mathbf{A} \approx -\{0, 1, 0\} ck \sin\theta \sin(kr) / r \, . \tag{A12}$$

Consequently each term of (A10) will have dimensions of $1/R$, so it is not obvious that the surface integrals will vanish at infinity. If the surface integrals (A10) are evaluated, they do in fact compute to zero, but this result is misleading as it relies upon the symmetry of the situation. If the cubic box is displaced from the origin by a vector $\alpha R\{1,1,1\}$, where $\alpha$ is a small number, it can be shown† that the surface integrals are changed by a term proportional to $\alpha^2 R$. As $R \rightarrow \infty$ the surface integrals diverge, because $\alpha$ is a finite (but small) number. Accordingly the decomposition (A8) is not valid in that case.

†A computer algebra script to verify these calculations is available from the author.